\documentclass[12pt]{article}

\usepackage[colorlinks=false,bookmarks=false,citecolor=black,linkcolor=black,urlcolor=black]{hyperref}

\usepackage{scicite2}

\usepackage[labelfont=bf,textfont=md]{caption}

\usepackage{times}
\usepackage{float}
\usepackage{lineno} 
\usepackage{siunitx} 
\usepackage{amsthm,amsmath,amssymb,amsfonts,bbm}
\usepackage{physics} 
\usepackage{graphicx} 
\usepackage[noabbrev]{cleveref} 

\Crefname{figure}{{Fig.}}{{Figs.}}
\crefname{figure}{{fig.}}{{figs.}}
\crefname{equation}{eqn.}{eqns.}%
\crefname{table}{table}{tables}
\Crefname{table}{Table}{Tables}

\title{Electromagnetic lensing using the Aharonov-Bohm effect} 

\author
{Makoto Tokoro Schreiber,$^{1,a,\ast}$ Cathal Cassidy,$^{2}$ Menour Saidani,$^{3,b}$ Matthias Wolf$^{1}$\\
	\\
	\normalsize{$^{1}$Molecular Cryo-Electron Microscopy Unit, Okinawa Institute of Science}\\
	\normalsize{and Technology Graduate University, 1919-1 Tancha, Onna-son, Okinawa, Japan}\\
	\normalsize{$^{2}$Quantum Wave Microscopy Unit, Okinawa Institute of Science and}\\
	\normalsize{Technology Graduate University, 1919-1 Tancha, Onna-son, Okinawa, Japan}\\
	\normalsize{$^{3}$Engineering Section, Research Support Division, Okinawa Institute of Science}\\
	\normalsize{and Technology Graduate University, 1919-1 Tancha, Onna-son, Okinawa, Japan}\\
	\normalsize{$^{a}$Present address: Department of Physics, University of Alberta, Edmonton, Canada}\\
	\normalsize{$^{b}$Present address: Nanofabrication Core Lab, King Abdullah University }\\
	\normalsize{of Science and Technology, Saudi Arabia}\\
	\\
	\normalsize{$^\ast$To whom correspondence should be addressed; E-mail:  mschreib@ualberta.ca}
}

\date{}

\begin{document}

	\baselineskip14pt	

	\maketitle

	\section*{Abstract:}
	We demonstrate theoretically and experimentally a new electromagnetic lensing concept using the magnetic vector potential – in a region free of classical electromagnetic fields – via the Aharonov-Bohm effect. This toroid-shaped lens with poloidal current flow allows for electromagnetic lensing which can be tuned to be convex or concave with a spherical aberration coefficient of opposite polarity to its focal length. This new lens combines the advantages of traditional electromagnetic and electrostatic field-based lenses and opens up new possibilities for the optical design of charged-particle systems. More generally, these results demonstrate that the Aharonov-Bohm effect can shape charged particle wavefronts beyond simple step shifts if topologies beyond simple flux lines are considered and supports the physical significance of the magnetic vector potential.

	\section*{One-Sentence Summary:} 
	The magnetic vector potential in the absence of magnetic fields can cause lensing effects for charged particles.

	\section*{Main text:}
	
	The concept of an electromagnetic lens was first introduced in 1926 by Hans Busch \cite{busch1926berechnung} and consists of a short solenoidal coil wrapping which produces a high magnetic field concentration in its central bore.
	Soon after this discovery, the first electron microscope was invented by Ruska and Knoll	\cite{ruska1987development} which opened up the ability to study a wide variety of structures, chemistry, and physics at length scales well beyond those of light microscopes.
	The lensing ability of these solenoidal electromagetic coils can be understood through the Lorentz force equation.
	The advantage of magnetic fields for lensing over electric fields is that the lensing forces are proportional to the electron energy as well as the field strength while electrostatic lensing is only proportional to the field strength.
	This allows a single electromagnetic lens to be used for a  wide range of electron energies.
	A major limitation of round electromagnetic lenses was discovered early on by Scherzer \cite{scherzer1936einige} who proved that these lenses will always be convex (positive focal length) and always have positive spherical ($C_s$) and chromatic ($C_c$) aberration  coefficients.
	The intrinsic aberrations of round electromagnetic lenses have long prevented direct electron imaging from reaching anywhere near wavelength-limited resolutions.
	Only in the last few decades has aberration correction with non-round multi-pole electromagnetic lens stacks become technologically possible \cite{krivanek1997line,haider1998electron,haider1998spherical,haider1998towards} and pushed the resolution limits of transmission electron microscopes (TEMs) to sub-angstrom ranges.
	
	Magnetic phenomena are usually explained in relation to the magnetic field. 
	In the 1950s, it was theorized by Aharonov and Bohm \cite{ehrenberg1949refractive,aharonov1959significance} that  electromagnetic potentials can cause physically observable phase shifts on charged particle waves in the absence of any electromagnetic fields.
	This was a non-intuitive result as the potentials are gauge fields and thus do not have a unique mathematical representation unlike  electromagnetic fields. 
	Due to this, the potentials in classical electrodynamics were widely considered to only be mathematical tools without direct physical significance.
	The absolute phase shifts induced by the Aharonov-Bohm (AB) effect on a charged particle cannot be directly observed.
	However, the relative phase shifts between charged particles passing through different regions of the electromagnetic potentials are gauge invariant and can be measured through interference experiments (as proposed in Aharonov and Bohm's original paper).
	Although both electrical scalar potentials and magnetic vector potentials were considered by Aharonov and Bohm (the magnetic effect also described earlier by Ehrenberg and Siday \cite{ehrenberg1949refractive}), the electrical version has not yet been satisfactorily verified experimentally \cite{batelaan2009aharonov}. 
	In this work, we only consider the magnetic AB effect.
	
	\section*{Phase shift profiles from special coil geometries}
	
	For a magnetic vector potential to exist in the absence of a magnetic field, non-simply connected geometries (space with ``holes'') are required.
	Two of the main geometries that have been considered are that of an infinite-length solenoidal coil and a toroidal solenoid coil; both with steady-state currents flowing along their surface.
	The infinite-length solenoid is the geometry originally considered by Aharonov and Bohm and has been the focus of more studies than the toroidal geometry.
	Early experimental tests of the AB effect used the solenoidal geometry in the form of magnetic whiskers \cite{chambers1960shift,fowler1961electron,bayh1962messung} or small solenoids \cite{mollenstedt1962kontinuierliche} but faced issues with magnetic field leakage due to the impossibility of using infinite-length objects.
	The widely accepted experimental verification of the magnetic AB effect came from Tonomura's group in the 1980s \cite{tonomura1986evidence,osakabe1986experimental} in which they measured the shift in interference fringes from electrons passing through a shielded toroidal magnet by electron holography.
	From calculations on infinite-length solenoids, it has been shown that particles traveling on the same side of the solenoid would experience no relative phase shifts \cite{kobe1979aharonov}.
	From this, a common assumption in the literature is that the AB phase shifts will only occur between particles whose paths enclose a non-zero magnetic field.
	We will demonstrate that this is not generally true and that relative phase shifts between trajectories enclosing no magnetic field can produce phase shift profiles with practical applications.
	
	We consider the cases of an infinite-length solenoidal coil and a toroidal solenoid coil with circular cross sections; both of which have analytic solutions for the vector potential in the Coulomb gauge \cite{carron1995fields}.
	Schematics of the coil geometries are shown in \Cref{fig:phase_profiles}A and B.
	We consider idealized versions of these coils such that steady-state currents flow with no helical pitch (\emph{i.e.} purely in the azimuthal $\hat{\phi}$ direction for the infinite-length solenoid and in the poloidal $\hat{\tau}$ direction for the toroidal solenoid).
	The relative phase shifts induced on electrons passing by each of these structures are calculated (details in \cref{sec:solenoid-theory,sec:torus-theory}).
	The shape of the integration loops used were chosen to approximate an electron plane-wave.
	\Cref{fig:phase_profiles}C shows a visual representation of the magnetic vector potential in the Coulomb gauge for the infinite-length solenoid.
	The relative phase shift profile for a plane-wave of electrons traveling in the positive $\hat{y}$ direction in \Cref{fig:phase_profiles}C is shown in \Cref{fig:phase_profiles}D.
	Here, it is apparent that there is a step-change in the relative phase from one side of the solenoid to the other (connected by a phase ramp for electrons passing through the region with magnetic field).
	If the current flow direction were flipped, the phase shifts would also be reversed.
	On the same side of the solenoid, there are no relative phase shifts as is generally expected for the AB effect.
	This lack of relative phase shift for particles passing on the same side of the infinite-length solenoid is a consequence of the vector field of the  magnetic vector potential outside the solenoid being conservative and thus path independent.
	While this aspect of the AB effect is usually ascribed to Stokes' theorem, Stokes' theorem cannot actually be applied here due to the magnetic vector potential being non-continuously differentiable when the whole system is considered.
	However, as the infinite-length solenoid in 3D is homotopy equivalent to a circle (current loop) in 2D, it is actually Green's theorem being applied where the flux within the solenoid does not have to be considered (\cref{sec:solenoid-theory}).

	The same calculations were repeated for the toroidal solenoid geometry as shown in \Cref{fig:phase_profiles}E-G.
	\Cref{fig:phase_profiles}E shows the analytic vector potential in the Coulomb gauge for the toroidal solenoid coil \cite{carron1995fields}. \Cref{fig:phase_profiles}F shows the calculated relative phase shift profile for an electron plane-wave traveling in the positive $\hat{z}$ direction.
	The phase shift profile of the torus exhibits a large step-shift in relative phase between regions outside the torus and inside the hole of the torus.
	This phase shift is what was observed in Tonomura's experiments.
	If the direction of the current through the torus is reversed, the phase shift profile is flipped.
	This is expected and was demonstrated in one of Tonomura's earlier experiments \cite{tonomura1982observation}.
	These features follow the traditional understanding of the AB effect.
	What is newly observed here is that the relative phase profile is not completely flat in the regions outside the torus volume.
	There is a small phase curvature of much lower magnitude than the well-known phase step-shift.
	This demonstrates that the vector potential outside the toroidal solenoid is non-conservative.
	As a consequence, the phase accumulated by charged particles passing through the regions with zero magnetic fields are no longer path-independent.
	Thus, the degree to which the phase profile curves will depend on the size and shape of the integration curves used (which is not the case with the infinite-length solenoid).
	
	The shape of the integration curves used for \Cref{fig:phase_profiles}E-G were chosen to approximate an electron plane-wave. 
	When we look more closely at the phase profile in the hole of the toroidal solenoid coil as shown in \Cref{fig:phase_profiles}G, the profile is close to parabolic which is a signature of lensing effects.
	An ideal lens phase profile can be fit very closely to the calculated relative phase profile in the torus.
	This suggests that the magnetic vector potential distribution inside the hole of a toroidal solenoid coil acts as a lens for charged particles.
	While this analysis precludes quantification of the lensing magnitude due to the path-dependent nature, some lensing properties that the toroidal solenoidal coil has can be identified.
	
	\section*{An electromagnetic toroidal solenoid coil lens}

	By changing the current flow through the toroidal solenoid coil, the lensing strength can be varied; just as for traditional solenoidal electromagnetic coils.
	In traditional electromagnetic lenses, electrons follow a conical spiral path through the lens. 
	By reversing the current flow direction, the handedness of the spiral reverses but not the focal length polarity.
	Therefore, these lenses can only be convex for negatively charged particles.
	With the toroidal solenoid coil, by reversing the current flow direction, the lensing direction of the coil can be flipped from convex to concave and vice-versa.
	Thus, the toroidal solenoid coil can act as a convergent or divergent lens for both positively and negatively charged particles (the focal length polarity will depend on the direction of current flow and the particle charge polarity).
	Additionally, no phase shifts are produced in the toroidal direction so no image rotation is produced as occurs with conventional electromagnetic lenses.
	In \Cref{fig:phase_profiles}G, we can see a reduced phase curvature at the  edge of the toroidal solenoid coil lens relative to that of an ideal lens.
	This indicates that the toroidal solenoid coil has a $C_s$ of opposite polarity to its focal length (\emph{e.g.} when the coil is operated such that it has a positive focal length, it has a negative $C_s$).
	This contrasts with a traditional electromagnetic lens which always has a positive non-zero $C_s$ for electrons.
	The $C_c$ of the toroidal solenoid coil will have the same polarity as the focal length due to the wavelength dependence of the beam deflection due to wavefront curvature (\cref{eq:deflection-angle}).
	
	That this toroidal solenoid coil lens has a spherical aberration coefficient of opposite polarity to its focal length is seemingly in contradiction with Scherzer's conditions \cite{scherzer1936einige,scherzer1949theoretical} for producing negative spherical aberration in electron lenses: 1. breaking rotational symmetry; 2. using time-varying fields; 3. using space-charges; and 4. producing a virtual image.
	The unstated assumption underlying these four conditions is that lensing takes place through electromagnetic fields.
	Scherzer's conditions thus do not apply to the present round, space-charge free, steady-state current lens as it is field-free.
	As the lens does not produce electromagnetic fields, it may be well suited for the imaging of magnetic samples.
	However, unlike the field-free objective lens recently developed by Shibata \emph{et al.} \cite{shibata2019atomic}, a magnetic vector potential will still be present at the sample. 
	The properties of this new lens are compared with existing lenses for charged particles in \cref{tab:comparison-table}.
	
	\section*{Further physics of the AB effect}
	
	Due to the path-dependent nature of the vector potential field, the charged particles passing through it should follow paths determined by stationary action principles.
	Using the classical electron-optical refractive index parameterized by distance along the optical ($z$)-axis, the Euler-Lagrange equation can be solved numerically to calculate the trajectory of charged particles passing through a toridal solenoid coil (details given in \cref{sec:lagrangian}).
	Examples of calculated ray trajectories for parallel electrons entering a toroidal solenoid coil are shown in \cref{fig:torus-ray-trajectories}.
	These trajectories exhibit the same lensing behavior predicted by the phase-based analysis; namely that the focal length and spherical aberration coefficients have opposite signs and that their polarities can be reversed by switching the direction of current flow.
	That this effect can be calculated using classical Lagrangian methods  challenges another previously held belief about the AB effect -- that it is a purely quantum mechanical effect.
	With the infinite-length solenoid geometry, deflections can only be observed through double-slit type interference experiments centered about the solenoid.
	Such deflections have been described as being due to a "quantum force" \cite{becker2019asymmetry}.
	The present lensing deflections for the toroidal geometry appear even in the classical limit were the wave-nature of the particle is not considered.
	Thus, while this lensing effect cannot be explained by the Lorentz force, aspects of it can be explained classically -- but only using the magnetic vector potential and not by the magnetic field.
	Therefore, the magnetic vector potential has physical significance not only in quantum mechanics but classically as well.

	In early critiques of the AB effect, it was argued that fringing magnetic fields from any physically achievable finite-length solenoid could also cause the shifts in the interference patterns as predicted by Aharonov and Bohm.
	However, as argued by Berry \cite{berry1986aharonov,berry1995three}, as the solenoid approaches longer and longer lengths, the phase shift must approach a non-zero asymptotic limit even as the fringing fields approach zero.
	For the present case of the toroidal solenoid system, it could be argued that it is not physically possible to achieve a perfectly axially symmetric torus.
	Such an imperfect torus could have fringing magnetic fields as shown in \cref{fig:torus-fringing-fields} which would give rise to roughly radial deflections similar to those predicted in this paper.
	As with the straight solenoid system, the current work argues that in the limit of vanishing fields (by improving the axial symmetry in this case), a non-vanishing radial lensing effect is maintained within the central hole of the torus.
	As the axial symmetry of the system is increased, the stray fields will be more closely confined to the edges of the torus.
	Experimentally, these edge-confined fringing fields (\emph{e.x.} in  atomically round samples) will prevent the sharp cusp in phase at the edges of the sample from being observable.
	As long as the observed lensing effect is radially symmetric, the contribution of stray fields would be negligible.

	It may be useful to note why previous quantum mechanical solutions to wave scattering from the toroidal magnetic field geometry \cite{ballesteros2009high,mine2018solvable} have not predicted the present lensing effects.
	This is because none of these previous theoretical calculation on the toroidal geometry have used magnetic vector potentials calculated directly from the toroidal solenoidal coil current distribution.
	Instead, the vector potential distributions were calculated from magnetic flux loops via Stokes' theorem. 
	In this work, we have shown that for geometries which deviate from the  infinite-length solenoid coil, application of Stokes' theorem is not valid (\cref{sec:stokes-breakdown}).
	Indeed, the infinite-length solenoid is a special case of the toroidal solenoid as the major radius $r_l$ becomes infinity.
	For the phase step-shift behaviors that previous quantum mechanical models were interested in justifying, the error induced by the application of Stokes' theorem is negligible.
	However, it hides the phase curvature effects which can only be found using the magnetic circulation (\cref{eq:magnetic-circulation}) rather than the magnetic flux.
	We expect that if scattering models are developed using explicit representations of the magnetic vector potential, wavefront curvatures through the torus will be found.
	Such quantum mechanical scattering-based solutions would allow for proper quantitative calculations of the lensing effects.

	The interpretation of the AB effect has been under debate for some time.
	In the original paper \cite{aharonov1959significance}, two interpretaions were offered; that the effect could be ascribed to non-local interactions between the charged particles and the magnetic fields or to local interactions between the charged particles and the magnetic vector potential.
	Subsequently, several papers have argued for each interpretation \cite{kasunic2019magnetic,aharonov2016nonlocality}.
	One of the main arguments for non-locality is that the AB effect only occurs when magnetic field lines (magnetic flux) is enclosed \cite{kobe1992asymmetry,wang2015possible,peshkin1989aharonov}.
	The current work demonstrates that phase shifts can occur between particles whose trajectories do not enclose any magnetic field.
	Thus, the current work further supports the physical significance of the magnetic vector potential.
	Any non-local interpretations will also need to account for the present phase curvature effect in non-straight solenoidal systems.
	
	\section*{Experimental demonstration of the lensing effect}
	
	To test the lensing theory, we fabricated magnetic nickel nanorings on silicon nitride membranes (\Cref{fig:diffraction_fresnel_expt}A and C).
	All experiments were performed in Lorentz-mode transmission electron microscopy (TEM) where the objective lens is off to minimize the effects of the magnetic fields from the (conventional solenoidal electromagnetic) lenses on the samples.
	As a non-magnetic reference, gold nanorings of similar size and thickness (\Cref{fig:diffraction_fresnel_expt}E) were also measured.
	Magnetic nanorings have several possible magnetization states.
	For the present experiment, we require toroidal magnetic fields within the ring material which is the flux closure magnetization state.
	Such rings were identified by electron holography (\cref{sec:holography}, \cref{fig:holograms}).
	The phase resolution of the holography measurements is however not sufficient to measure the phase curvature.
	
	To measure the lensing effect from magnetic nickel nanorings with flux-closure vortex magnetization states, convergent nanobeams smaller than the central hole of the nanorings were formed.
	The probes were imaged in diffraction mode defocused away from the back focal plane such that Fresnel diffraction patterns were recorded as the electron beam passed through the ring.
	Reference diffraction patterns were obtained in vacuum without changing the beam parameters.
	In \Cref{fig:diffraction_fresnel_expt}B, the Fresnel fringes from a beam passing through the hole of a nickel nanoring are shifted slightly radially outwards relative to the reference beam through vacuum as would be expected for a beam being diverged (\cref{fig:fresnel_diff_lens_sim}).
	The nanoring grid was then physically flipped to reverse the magnetization direction relative to the electron beam and Fresnel patterns collected from the same ring are shown in \Cref{fig:diffraction_fresnel_expt}D.
	Here, the Fresnel fringes are observed to shift slightly radially inwards as expected for convergent lensing.
	The non-magnetic gold nanoring control samples (\Cref{fig:diffraction_fresnel_expt}E) showed no shift in the Fresnel fringe patterns between the inside of the nanorings and vacuum (\Cref{fig:diffraction_fresnel_expt}F).
	The lensing polarities are consistent with what our theory (\Cref{fig:phase_profiles}F and G) would predict based on the phase profiles measured by electron holography through the whole nanoring structures (\cref{fig:holograms}).
	
	The samples used were not magnetically shielded or electron opaque as in the experiments performed by Tonomura's group \cite{tonomura1986evidence,osakabe1986experimental}.
	This allowed for holography to be performed on the whole structure but means that some magnetic field may extend into the hole of the nanorings.
	While reversible concave-convex lensing is possible using time-dependant electromagnetic fields in intense laser pulses \cite{uesugi2021electron,zhang2018segmented,constantin2022transverse},
	as proved by Scherzer, there are no time-independent magnetic fields which can extend into free-space and produce radially symmetric convex and concave reversible phase profiles as observed in the present Fresnel diffraction experiments.
	If the currently observed fringe shifts were due to electrons passing through the volume of the nanorings (a magnetic field-based lensing effect \cite{harris1947proposed,horvath1949notices,mohri1977new}), the lensing effects would be in the opposite directions from those presently measured.
	Beam-induced charging effects of the support membrane also cannot explain the observed shifts in Fresnel fringes as the polarity of any induced charge would be the same whether the sample were flipped upside down or not and thus not reverse the fringe-shift direction.
	Thus the present Fresnel diffraction experiments provide experimental support for our theoretical conclusions.
	
	To further establish the lensing effect, some other improvements in addition to  shielded and flux confining nanorings can be considered.
	Even with magnetic shielding around the sample, there is a non-zero possibility for electrons passing through the hole of the nanoring to interfere with electrons passing outside the nanoring -- leading to Shelenkov-Berry \cite{shelankov1998magnetic,berry1999aharonov,becker2016experimental} type deflections.
	To eliminate such possibilities, a thick metallic sheet with a hole smaller than the inner hole of the magnetic nanoring could be placed over the nanoring.
	This would prevent even the weak tails of a convergent electron probe from extending beyond the inner hole to truly verify that deflections can be caused without enclosing any magnetic flux.
	Additionally, if experiments could be performed using massive charged particles with negligible deBroglie wavelength, the ``classical'' aspect of the lensing effect could also be verified.
	
\begin{figure}
		\centering
		\includegraphics[width=1\linewidth]{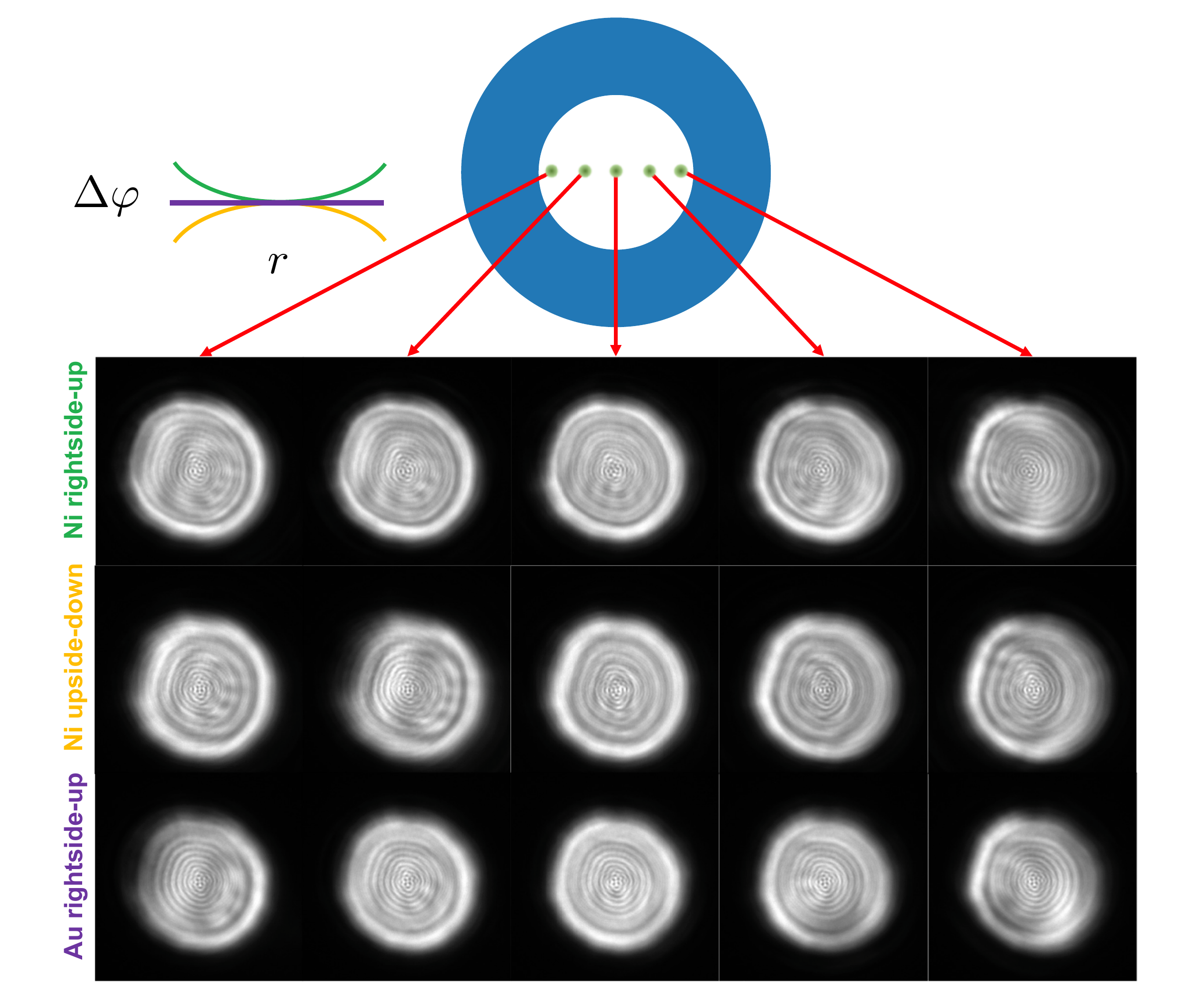}
		\caption{\textbf{Fresnel diffraction patterns obtained at different radial positions through the center of nanorings} (a magnetic nickel nanoring in the rightside-up and upside-down orientations and a non-magnetic gold nanoring).
			The top-center schematic represents the nanoring (blue ring) and the approximate positions (green dots) of the electron beam from which the corresponding diffraction patterns were obtained (columns under red arrows). 
			The top-left diagram represents the shape of the phase profiles expected through the three rings shown (green concave = Ni rightside-up, yellow convex = Ni upside-down, purple flat = Au).  
			The Fresnel fringes in the diffraction patterns obtained off-center through the magnetic nanoring exhibit left-right asymmetry. 
			The off-center obtained Fresnel diffraction patterns through the non-magnetic nanorings do not exhibit left-right asymmetry.}
		\label{fig:diff-off-center}
	\end{figure}

	As an additional test of the lensing effect due to the magnetic vector potential with the current sample, we also collected Fresnel diffraction patterns from convergent electron beams not centered in the nanorings (\emph{i.e.} not along the optic axis).
	For axially symmetric lenses, radiation passing off the optic axis will gain a non-axially symmetric phase shift.
	A series of Fresnel diffraction patterns were obtained after a convergent electron probe was formed and the sample stage translated along one direction such that the electron probe was scanned through the center of a nanoring as shown in the top diagram of \Cref{fig:diff-off-center}.
	As before, the microscope lens settings were not modified within each scan.
	Due to lack of precision in the stage shifts and drift, the exact position of the electron probe in the nanoring cannot be determined but the approximate positions are as depicted in the diagram.
	For the magnetic nanorings (the same ring as used in previous experiments), when the electron probe passes near the inner edge of the nanoring, a left-right asymmetry in the Fresnel fringe spacings is observed.
	As the probe is moved to the center of the nanoring, the asymmetry is no longer observed (the same conditions as in \Cref{fig:diffraction_fresnel_expt}).
	As the probe continues to the other side of the nanoring, the asymmetry reappears but with the opposite orientation.
	For the electron probe passing through the non-magnetic gold nanoring, such asymmetry in the Fresnel fringes is not observed. 
	Some image contrast in addition to the Fresnel fringes is observed in all cases due to irregularities in the support film (partially due to repeated exposures to the central region of the ring).
	This contrast can be identified as it moves with the nanoring.
	When the magnetic nanoring is flipped upside-down, the way the left-right asymmetry in the fringe spacings appear reverses.
	For the right-side up case in \Cref{fig:diff-off-center}, fringes toward the center of the nanoring are more closely spaced relative to the fringes closer to the edge of the nanoring.
	When the nanoring is upside-down, the fringes towards the center now have a wider spacing relative to the fringes toward the edge of the nanoring.
	These distortions to the fringes are again consistent with the results of the previous experiments and theory.
	We have thus experimentally demonstrated both on-axis and off-axis lensing effects of the magnetic vector potential from toroidal magnetic structures.

	\section*{Conclusion}
	
	We have argued theoretically and demonstrated experimentally that the irrotational magnetic vector potential in the bore of a toroidal solenoid coil can act as an electromagnetic lens for charged particles through the Aharonov-Bohm effect. 
	This electromagnetic lens is unique compared to existing round, time-independent, space-charge-free lenses as it can be made both divergent and convergent and has a negative spherical aberration coefficient while convergent.
	This new lensing concept has potential to be practically applied in charged particle optics systems such as electron microscopes.
	The aberration and reversable focal length properties in particular make it theoretically possible to design microscopes with near-zero spherical aberrations without the need for multi-pole aberration correctors.
	This lensing also for the first time demonstrates an Aharonov-Bohm effect where the measurable relative phase shift does not take place between particle trajectories enclosing a magnetic field.
	This supports the interpretation of a localized interaction between the charged particle and magnetic vector potential.
	Additional effects of the magnetic vector potential and perhaps other gauge fields may be found if higher order topological structures are investigated.

	\renewcommand\refname{References:}
	\bibliography{references}
	
	\bibliographystyle{sciencetitle}

	\section*{Acknowledgments:}

	We would like to thank Dr. Martin Linck (CEOS GmbH) for providing a protocol for the CETCOR image corrector to allow for larger fields of view in electron holography. 
	We would like to thank Professor Takuya Mine (Kyoto Institute of Technology), Professor Dmitry Feichtner-Kozlov (University of Bremen, OIST), and Professor Rafael Ayala (University of Seville) for helpful discussions.
	The OIST Imaging Section is acknowledged for use of the ETEM facility. The OIST Engineering Section is acknowledged for use of the nanofabrication facilities.
	
	\subsection*{Funding:}
	M.T.S. was supported by JSPS fellowship (DC1) 201820215, JSPS Kakenhi grant 18J20215 and Kakenhi Grant-in-aid for JSPS fellows 18J20132. 
	C.C. was supported by JSPS Kakenhi Grant No. 18KO4247.
	M.W. was supported by the Platform Project for Supporting Drug Discovery and Life Science Research (BINDS) from AMED under grant number JP18am0101076. 
	M.T.S, C.C and M.W. are grateful for direct funding from OIST.
	
	\subsection*{Author contributions:}
	M.T.S. conceived of the study and conducted the theoretical calculations.
	M.S. fabricated the nanoring samples. 
	M.T.S. and C.C. performed the electron microscopy experiments and data analysis.
	M.T.S. wrote the manuscript.
	M.W. provided supervision. 
	
	\subsection*{Competing interests:}
	A patent application on the toroidal solenoid electromagnetic coil lens has been filed by OIST with M.T.S., C.C., and M.W as inventors.
	
	\subsection*{Data availability:}
	The data from this study is available upon request.
	
	\begin{figure*}
		\centering
		\includegraphics[width=\linewidth]{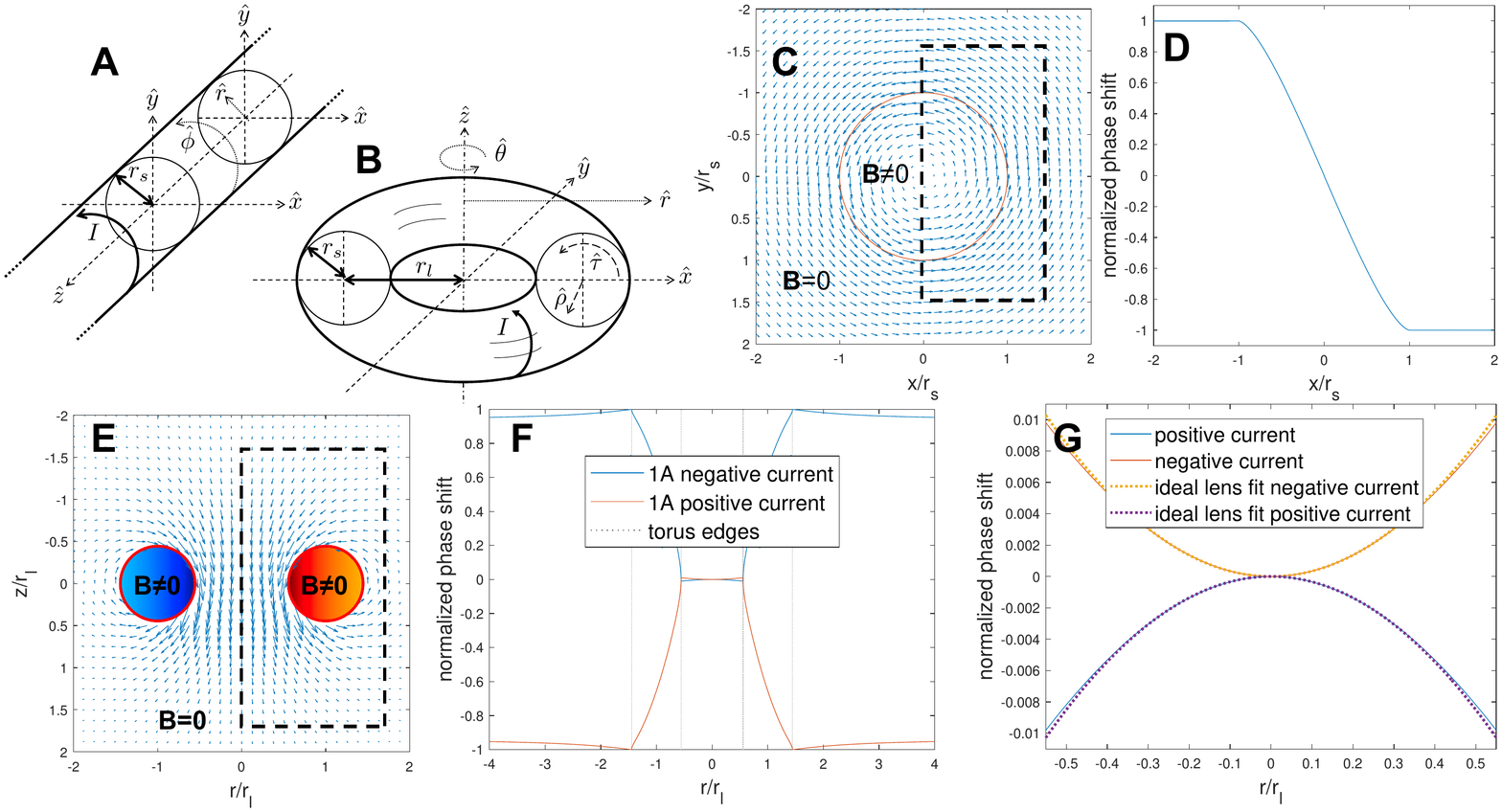}
		\caption{\textbf{Calculations of the relative phase shift profiles from two coil geometries.} 
			(\textbf{A}) Schematic diagram of the infinite-length solenoidal coil with circular planar cross-section showing coordinate directions (Cartesian $(x,y,z)$ and cylindrical $(r,\phi,z)$), shape parameter radius $r_s$, and current flow $I$ in the azimuthal $\hat{\phi}$ direction. 
			(\textbf{B}) Schematic of the toroidal solenoid coil with circular planar and axial cross-sections showing coordinate directions (Cartesian $(x,y,z)$, cylindrical $(r,\theta,z)$, and simple toroidal $(\rho,\theta,\tau)$), shape parameters minor radius $r_s$ and major radius $r_l$, and current flow in the poloidal $\hat{\tau}$ direction. Here, the magnetic field inside the torus volume would flow in the negative toroidal $\hat{\theta}$ direction. 
			(\textbf{C}) Vector plot of the magnetic vector potential (blue arrows) in the Coulomb gauge for a planar cross-section through an infinitely-long solenoidal coil. The red circle indicates the walls of the solenoid. The dashed rectangle represents an example integration loop. 
			(\textbf{D}) Phase profile for an electron plane-wave that has traveled perpendicular to the solenoid in the positive $\hat{y}$ direction. The phase shifts are relative to an electron traveling along the $y$ axis. 
			(\textbf{E}) Vector plot of the magnetic vector potential in the Coulomb gauge (blue arrows) for an axial cross section of a toroidal solenoid coil. The color scale indicates the magnetic field strength (blue going into the page, red coming out of the page, white=zero). The red circles indicate the walls of the torus. The dashed rectangle represents an example integration loop. 
			(\textbf{F}) Phase profile for an electron plane-wave traveling in the positive $\hat{z}$ direction that has passed through the toroidal solenoid coil. The phase shifts are relative to an electron traveling along the $z$ axis. The red line indicates the phase profile for a positive current and the blue line for a negative current. The dotted vertical lines represent the boundaries of the toroidal coil. 
			(\textbf{G}) The relative phase profile from (\textbf{F}) zoomed into the central hole of the toroidal coil. The dotted lines represent fits of an ideal lens to the phase profiles in the hole of the toroidal solenoid coil. 
			The phase profile inside the hole of a toroidal solenoid coil suggests that the magnetic vector potential in this geometry can focus or diverge charged particles.
		}
		\label{fig:phase_profiles}
	\end{figure*}

	\begin{figure*}
		\centering
		\includegraphics[width=\linewidth]{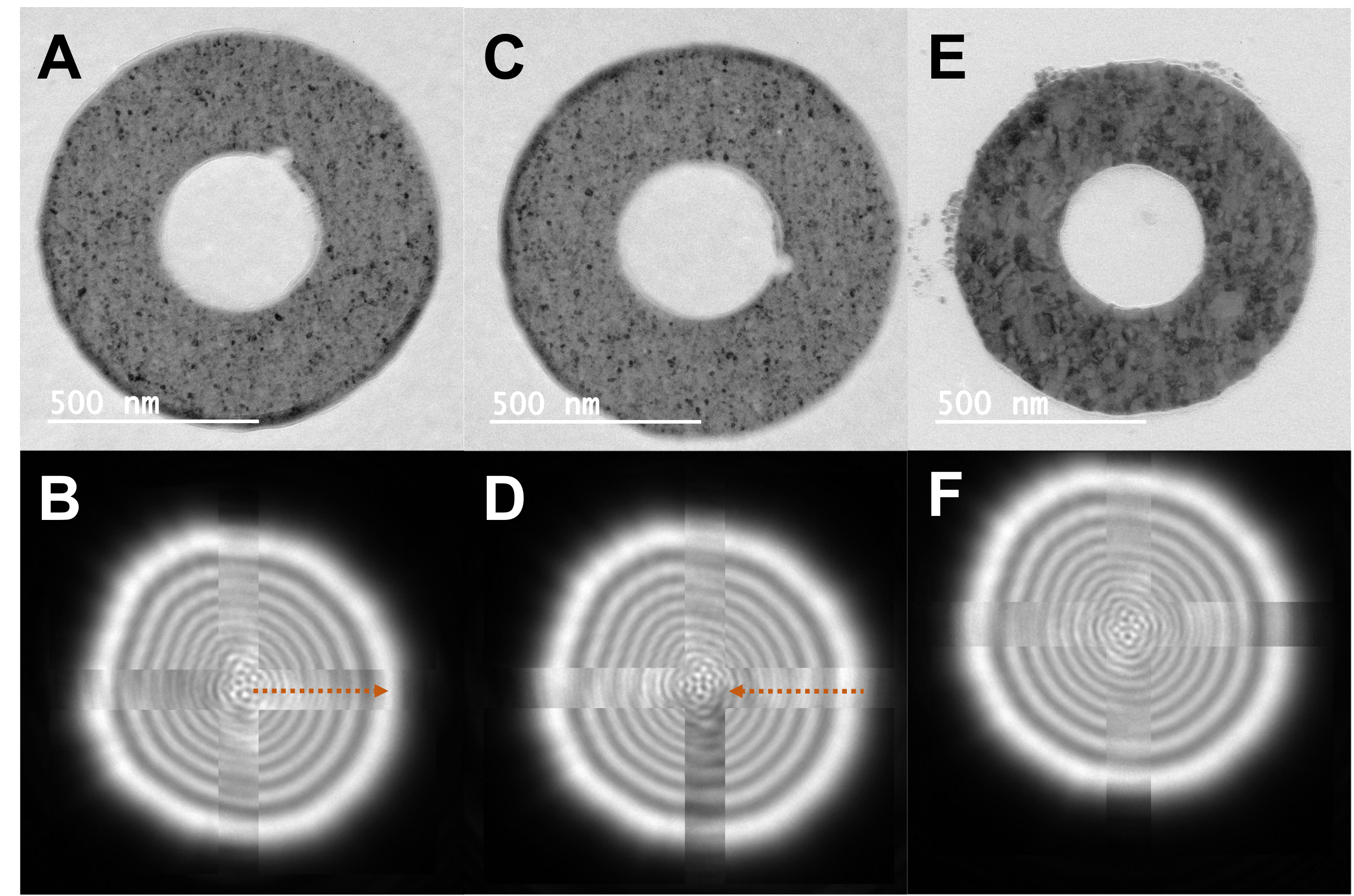}
		\caption{\textbf{Experimental demonstration of the lensing phase profile inside a magnetic nanoring.}
			(\textbf{A}, \textbf{C}, \textbf{E}) Lorentz TEM images of nanoring structures. 
			(\textbf{B}, \textbf{D}, \textbf{F}) Convergent beam Fresnel electron diffraction patterns obtained with the beam inside the nanorings and in vacuum. The beam conditions were unchanged while the sample was moved for acquisitions in the nanorings and in vacuum. 
			In these images, the Fresnel patterns from inside the nanorings and from vacuum are overlayed.
			The outer quadrants of the beam are the vacuum measurements while the inner cross is from inside the nanoring.
			(\textbf{A}, \textbf{B}) Nickel magnetic nanoring. 
			(\textbf{C}, \textbf{D}) the same nanoring as in (\textbf{A}, \textbf{B}) but physically flipped to reverse the magnetization direction relative to the electron beam. 
			(\textbf{E}, \textbf{F}) Gold non-magnetic nanoring. For the non-magnetic nanoring, there is no change in the Fresnel fringe pattern inside the nanoring and in vacuum. For the magnetic nanorings, the Fresnel pattern obtained inside the ring shifts relative to the vacuum pattern. The direction that the fringes shift reverse when the magnetic nanoring is flipped upside-down. The fringe shift directions are indicated with dotted arrows.}
		\label{fig:diffraction_fresnel_expt}
	\end{figure*}
	
	\pagebreak

	\setcounter{section}{0}
	\renewcommand{\thesection}{S\arabic{section}}
	\setcounter{figure}{0}
	\renewcommand{\thefigure}{{S\arabic{figure}}}
	\setcounter{table}{0}
	\renewcommand{\thetable}{{S\arabic{table}}}
	\setcounter{equation}{0}
	\renewcommand{\theequation}{S\arabic{equation}}

	\setcounter{page}{1}
	\setcounter{linenumber}{1}
	\section*{}
	
	\begin{center}
		\LARGE \textbf{Supplementary materials for}
	\end{center}
	\begin{center}
		\Large \textbf{Electromagnetic lensing using the Aharonov-Bohm effect}
	\end{center}
	
	\begin{center}
		Makoto Tokoro Schreiber, Cathal Cassidy, Menour Saidani, Matthias Wolf\\
		Correspondence to:  mschreib@ualberta.ca, mwolf@oist.jp
	\end{center}

	\subsection*{This PDF file includes:}
	Materials and Methods\\
	Supplementary Text\\
	Figs. S1 to S5\\
	Table S1\\
	References \textit{(41-49)}

	\section{Methods and Materials}
	
	\subsection{Vector potential phase shift profile calculations}
	
	Calculations were performed in MATLAB R2020b.
	Vector potentials fields in the Coulomb gauge were calculated based on analytical solutions for given geometrical and current parameters.
	The elliptic integrals were evaluated with the built-in function `ellipke' with the default tolerance value with double floating point precision.
	Phase profiles were calculated by numerical integration of rectangular closed loops through the vector potential fields.
	
	The geometric particle trajectories were calculated from using the NDSolve function in Mathematica 12.
	
	\subsection{Nanoring fabrication}
	Magnetic nickel nanorings of 60-\SI{70}{\nano\meter} thickness with a \SI{5}{\nano\meter} titanium adhesion layer were grown on \SI{8}{\nano\meter} thickness silicon nitride membranes (Electron Microscopy Sciences).
	Non-magnetic gold nanorings of 60-\SI{70}{\nano\meter} thickness with a \SI{5}{\nano\meter} titanium adhesion layer were grown on \SI{50}{\nano\meter} thickness silicon nitride membranes (Norcada).
	Silicon nitride TEM window grids were cleaned for 5 min. with a soft oxygen plasma.
	\SI{300}{\nano\meter} poly(methyl methacrylate) (PMMA) 950K A4 was spin-coated onto the membranes and then soft baked for 3 minutes for solvent release.
	An ELS-7500EX electron-beam lithography system (Elionix, \SI{50}{\kilo\volt} acceleration, \SI{100}{\pico\ampere} current, \SI{30}{\micro\meter} aperture) was used to write an array of nanoring designs with 200-\SI{600}{\nano\meter} inner diameters and 500-\SI{1000}{\nano\meter} outer diameters.
	The exposed layer of PMMA was then developed in MIBK:IPA solution at 1:3 ratio and consequently rinsed in isopropyl alcohol (IPA).
	A KE604TT1-TKF1 electron-beam evaporation system (Kawasaki Science) was used to deposit metals onto the exposed PMMA films.
	PMMA lift-off was achieved by soaking the grids for 1-2 hours in acetone followed by a pipette flush with acetone and then in IPA.
	
	The grids were subsequently coated with $\sim$\SI{10}{\nano\meter} amorphous carbon using a IB-29510VET vacuum evaporator (JEOL) to reduce charging of the silicon nitride film during TEM observation.
	
	\subsection{Electron Microscopy}
	
	A Titan G2 ETEM (Thermo Fisher Scientific) operated at 300 kV was used for holography and diffraction experiments on the nanorings. 
	The microscope has a monochromator, Schottky XFEG electron source, S-TWIN objective lens, Lorentz lens, image Cs-corrector (CEOS GmbH), a single electrostatic biprism in the selected area plane, and a post-column Gatan Quantum 966 energy filter.
	UltrascanXP1000 (Gatan) cameras are located before and after the energy filter.
	Samples were loaded onto a model 2020 single-tilt tomography holder (Fischione).
	All experiments were performed in Lorentz mode such that the objective lens was off -- minimizing the magnetic field at the sample plane.
	
	Electron holography was performed on the nanoring samples to determine the magnetization states.
	As our single-biprism system has a limited field of view, a  protocol from CEOS was used to manipulate the lenses in the image corrector to reduce the image magnification transferred to the selected area plane.
	Holograms were obtained from nanorings with the grid right-side up and upside-down to observe the reversal of the magnetization state with respect to the electron beam.
	The hologram reconstructions from the right-side up and upside-down nanorings were subtracted to isolate the magnetic contributions to the hologram.
	Reference holograms were obtained from substrate films areas far from the nanorings.
	A biprism bias of \SI{100}{V} and camera exposure time of \SI{2}{\second} was used and a fringe contrast of $>14\%$ obtained on the substrate.
	The holograms were reconstructed using a custom script in MATLAB with a 50 pixel radius hard virtual aperture.
	A Magnitude-Sorted List, Multi-Clustering Phase Unwrapping Algorithm \cite{maier2015robust} was utilized to unwrap the phase maps.
	The phase maps were smoothed with a 16 pixel Gaussian blur filter to reduce the influence of a few regions which could not be properly phase unwrapped.
	
	For the Fresnel diffraction patterns measurements, the electron beam was converged to a size smaller than the inner radius of the nanorings ($<$ \SI{100}{\nano\meter}).
	The electron beam was centered in the nanoring and Fresnel diffraction patterns recorded.
	After recording the Fresnel pattern from the beam in the hole of the nanoring, a reference pattern was recorded in vacuum by moving the sample so that the beam conditions were unchanged.
	Diffraction pattern pairs were obtained for nanorings both in the right-side up and upside-down orientations.
	
	\section{Theory}
	\label{sec:phase-calcs}

	In wave optics, lenses are objects which impart spatially dependent phase shifts upon a plane wave in such a way that the waves will be focused to a point (limited in size by the radiation wavelength) after a given propagation distance (the focal length $f$ of the lens).
	For an ideal round lens, the phase shift imparted to an incoming wave is parabolic
	\begin{linenomath*}
		\begin{equation}
			\label{eq:ideal-lens}
			\Delta \varphi_\text{ideal} = - \frac{\pi r^2}{\lambda f}
		\end{equation}
	\end{linenomath*}
	where $r$ is the radial distance from the optical axis of the lens and $\lambda$ is the wavelength of radiation passing through the lens.
	
	For the case of charged particle optics -- or electron-optics in the case of electron microscopy -- electromagnetic or electrostatic lenses are used to influence the trajectories of the charged particles.
	The lensing effects can be understood by the Lorentz force 
	\begin{linenomath*}
		\begin{equation}
			\boldsymbol{F} = q (\boldsymbol{E} + \boldsymbol{v} \times \boldsymbol{B})
		\end{equation}
	\end{linenomath*}
	where $q$ is the electric charge of the particle with velocity $\boldsymbol{v}$, $\boldsymbol{E}$ is the electric field, and $\boldsymbol{B}$ is the magnetic field.
	For high-energy particles such as those used in a typical transmission electron microscope (TEM), electromagnetic lenses are preferred as the lensing force is proportional not only to the field strength but also to the particle velocity (which is proportional to the particle energy).
	Electrostatic lenses are not practically useful for such high-energy particles but are used for lower particle energies such as in scanning electron microscopes (SEMs).

	For typical electromagetic lenses, it is considered that only a $\boldsymbol{B}$ field and no $\boldsymbol{E}$ field is present.
	For a magnetic field, we can define a magnetic vector potential $\boldsymbol{A}$ similar to the scalar electric potential $V$.
	It is related to the magnetic field by
	\begin{linenomath*}
		\begin{equation}
			\curl{\boldsymbol{A}} = \boldsymbol{B}.
		\end{equation}
	\end{linenomath*}
	$\boldsymbol{A}$ is a gauge field and is thus not uniquely defined as any number of vector potential representations are possible through the gauge transformation $\boldsymbol{A'} = \boldsymbol{A} + \grad{\chi}$ where $\chi$ is a twice continuously differentiable scalar function.
	For convenience, in the case of steady-state currents (magneto-statics), we choose to represent $\boldsymbol{A}$ in the Coulomb gauge where we require 
	\begin{linenomath*}
		\begin{equation}
			\div{\boldsymbol{A}} = 0.
		\end{equation}
	\end{linenomath*}
	Calculations from gauge fields which do not depend on a specific gauge representation are known as gauge invariants.
	All physically observable (measurable) properties are gauge invariant.

	For a charged particle passing through a magnetic field, the phase shift it experiences relative to a reference trajectory can be calculated based on the magnetic flux $\Phi_B$ enclosed by the two trajectories
	\begin{linenomath*}
		\begin{equation}
			\label{eq:phase-flux}
			\Delta \varphi = \frac{q}{\hbar}\Phi_B
		\end{equation}
	\end{linenomath*}
	where $\hbar$ is the reduced Plank's constant.
	The magnetic flux is defined as the integrated perpendicular component of the magnetic field passing through an enclosed surface $\boldsymbol{S}$.
	By Stokes' theorem, this can be related to the line integral of the magnetic vector potential along the boundary $\partial S$ of the enclosed surface $S$
	\begin{linenomath*}
		\begin{equation}
			\label{eq:flux}
			\Phi_B \equiv \iint_S \boldsymbol{B}\cdot\text{d}\boldsymbol{S} = \oint_{\partial S} \boldsymbol{A}\cdot\text{d}\boldsymbol{l}
		\end{equation}
	\end{linenomath*}
	where $\text{d}\boldsymbol{l}$ is a line element.
	The second equality only strictly holds when considering a simply-connected region (a topology in which  all closed loops within the domain can be continuously transformed to a single point).
	Another way of stating this is that $\boldsymbol{A}$ must be continuously differentiable everywhere.
	
\begin{figure}[p]
		\centering
		\includegraphics[width=0.4\linewidth]{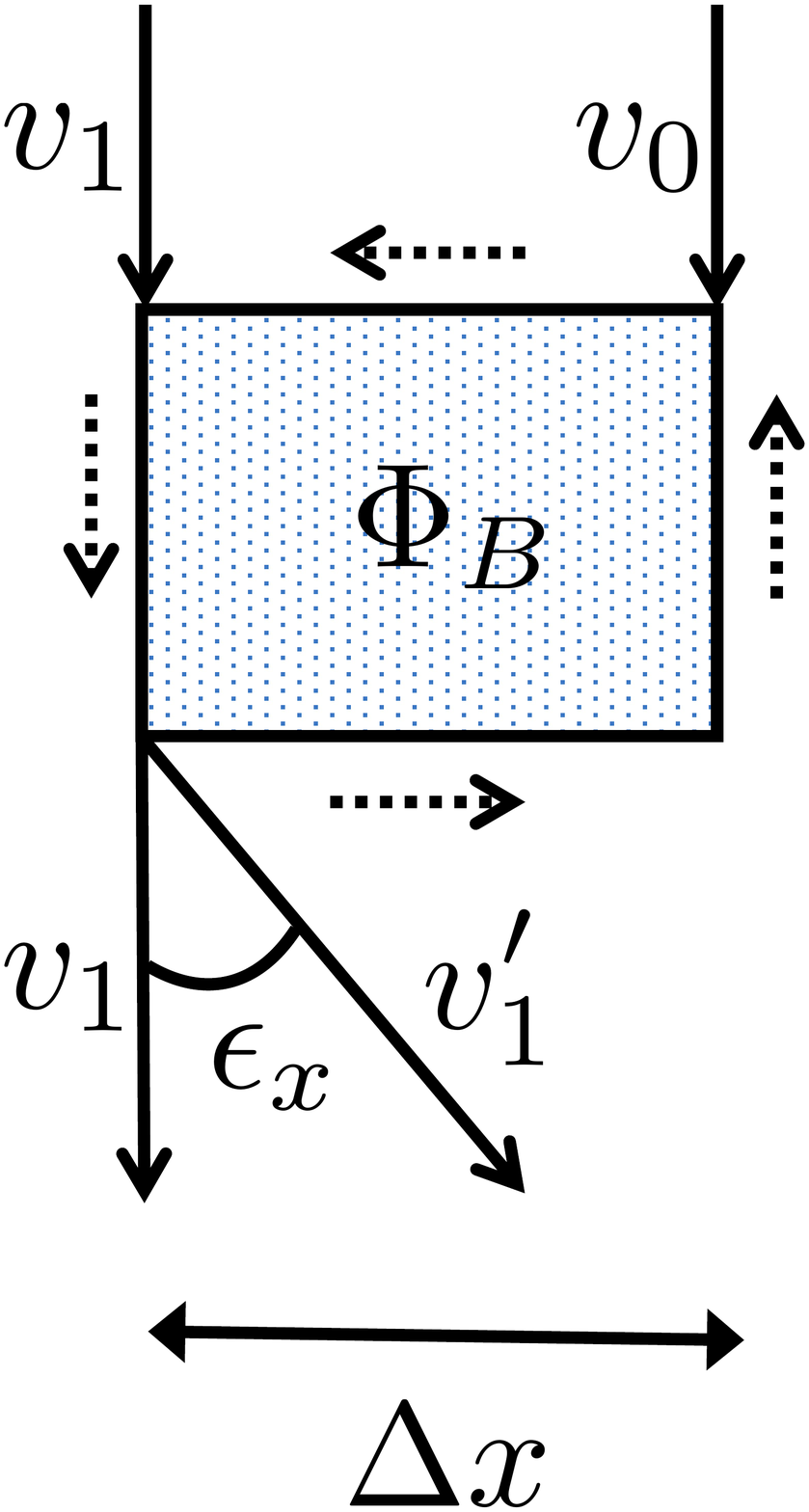}
		\caption{\textbf{Schematic of the deflection angle in relation to the accumulated phase shift.} The box represents the enclosed magnetic flux used to calculate the phase shift. Inside the box, a uniform magnetic field is coming out of the page. The dotted arrows outside the box represent the magnetic vector potential circulation direction in the Coulomb gauge. $v_1$ represents the electron trajectory of interest and $v_0$ the reference electron trajectory. $v_1'$ represents the trajectory of interest after undergoing a phase shift.}
		\label{fig:deflection-angle}
	\end{figure}

	If a non-simply connected geometry is considered, a region of space with zero magnetic field and non-zero magnetic vector potential can be generated.
	The relative phase shifts induced in particles traveling through such regions is known as the (magnetic) Aharonov-Bohm (AB) effect \cite{aharonov1959significance,ehrenberg1949refractive}. 
	For electrons with charge $q_e=$\SI{-1.602e-19}{C}, the effect is 
	\begin{linenomath*}
		\begin{equation}
			\label{eq:AB}
			\Delta \varphi_\text{AB} = \frac{q_e}{\hbar} \oint \boldsymbol{A}\cdot \text{d} \boldsymbol{l}.
		\end{equation}
	\end{linenomath*}
	and is gauge-invariant \cite{babiker1984gauge}.
	Note that while the AB effect specifically refers to regions with zero magnetic field, the phase shift equation also applies to regions with non-zero magnetic fields.
	The phase shift can be related to the deflection angle of the beam by \cite{reimer2013transmission}
	\begin{linenomath*}
		\begin{equation}
			\label{eq:deflection-angle}
			\epsilon_x = \frac{\lambda \Delta \varphi_x}{2 \pi \Delta x}
		\end{equation}
	\end{linenomath*}
	where $x$ is the direction along which the phase shift is calculated and $\Delta x$ is the length over which the phase shift was calculated in the $x$ direction relative to a reference trajectory (\cref{fig:deflection-angle}).

	In the following sections, we will explore the AB-effect phase shift profiles for two geometries: an infinite-length solenoidal coil and a toroidal solenoid coil. 
	
	\subsection{An infinite-length solenoidal coil}
	\label{sec:solenoid-theory}
	
	For a circular cross-section infinite-length solenoidal coil of radius $r_s$ with current flowing purely in the azimuthal $\hat{\phi}$ direction (main text \Cref{fig:phase_profiles}A), the $\boldsymbol{B}$-field is zero outside the solenoid and exists only in the $\hat{z}$ direction inside the solenoid with a constant value $B_0=\mu_r \mu_0 N_L I_s$.
	$I_s$ is the current through the wires, $N_L$ is the number of turns per unit length, $\mu_0$ is the permeability of free space, and $\mu_r $ is the relative magnetic permeability of the core material inside the solenoid.
	The magnetic vector potential of the infinite-length solenoid coil $\boldsymbol{A_{is}}$ in the Coulomb gauge is purely in the $\hat{\phi}$ direction and exists both inside and outside the solenoid.
	In cylindrical coordinates, 
	\begin{linenomath*}
		\begin{equation}
			\label{eq:infinite-solenoid}
			\boldsymbol{A_{is}}(r,\phi,z) = 
			\begin{cases}
				\frac{\mu_r  \mu_0 N_L I_s r_s^2}{2 r} \hat{\phi}, & (r>r_s) \\
				\frac{\mu_r \mu_0 N_L I_s r}{2} \hat{\phi}, & (r<r_s)
			\end{cases}
		\end{equation}
	\end{linenomath*}
	where $r$ is the radial distance defined perpendicular to the $z$-axis.
	A plot of $\boldsymbol{A_{is}}$ is shown in main text \Cref{fig:phase_profiles}C.
	
	For the electrons traveling in the $\hat{y}$ ($(x_1,y_0)\rightarrow(x_1,y_1)$) direction, the phase shift profile relative to another electron traveling in the same direction ($(x_0,y_0)\rightarrow(x_0,y_1)$) can be calculated by
	\begin{linenomath*}
		\begin{multline}
			\label{eq:loop-integral-solenoid}
			\frac{\hbar}{q_e}\Delta \varphi_x(x_1,y_0 \rightarrow y_1) = \oint \boldsymbol{A_{is}}\cdot\text{d}\boldsymbol{l'} \\ =  \int_{x_0}^{x_1}A_{is,x}(x,y_1) \text{d}x - \int_{x_0}^{x_1} A_{is,x}(x,y_0) \text{d}x  \\ + \int_{y_0}^{y_1} A_{is,y}(x_0,y) \text{d}y - \int_{y_0}^{y_1} A_{is,y}(x_1,y) \text{d}y.
		\end{multline}
	\end{linenomath*}
	A calculated phase profile with respect to electrons traveling along the $y$-axis ($x_0=0$) is shown in main text \Cref{fig:phase_profiles}D.

	The vector field $\boldsymbol{A_{is}}$ is conservative in any sub-region outside the infinite-length solenoidal coil.
	This is because the vector potential can be written as a gradient of a scalar function $\boldsymbol{A_{is}}(r>r_s)=\grad{\psi}$ where
	\begin{linenomath*}
		\begin{equation}
			\label{eq:solenoid-scalar}
			\psi(r>r_s,\phi,z) = \frac{\mu_r  \mu_0 N_L I_s r_s^2}{2}\phi.
		\end{equation}
	\end{linenomath*}
	Thus necessarily the vector potential is irrotational (zero curl) and path independent.
	However as a whole, the vector field of the solenoid is not conservative.
	It is not a general statement that magnetic vector potentials in sub-regions with zero magnetic field are conservative for non-simply connected geometries.
	A non-conservative case and its consequences will be demonstrated with the toroidal solenoid coil geometry in the next section.
	
	The conservative nature of the vector field outside the infinite-length solenoid is a consequence of the topology of the system.
	The topology is that of a 2D cylinder embedded in a 3D space $\mathbb{R}^3 \setminus (S^1 \times (0,1))$.
	This topology is homotopy equivalent to a circle embedded in a 2D space $\mathbb{R}^2 \setminus S^1$ \footnote{Even stronger, the circle is a strong deformation retract of the cylinder.}.
	The circle $S^1$ can also be referred to as a 1D sphere (1-sphere) \footnote{An n-sphere is defined as $S^n \equiv \left\{(x_0,\dots, x_n) \in \mathbb{R}^n, x_0^2+ \dots + x_n^2=1 \right\}$}
	or 1D torus (1-torus) \footnote{An n-torus is the Cartesian product of $n$ 1-spheres $T^n \equiv \underbrace{S^1 \times \dots \times S^1}_n$}.
	Therefore, applying Stokes' theorem to the vector field around an infinite-length solenoid coil is equivalent to applying Green's theorem to the 2D vector field around a current loop.
	In Green's theorem, the presence of holes within an enclosed area can be dealt with by adding the circulations along the boundaries of the holes
	\begin{linenomath*}
		\begin{equation}
			\sum_i \oint_{\partial S_i} \boldsymbol{A} \cdot \text{d}\boldsymbol{l}  = \iint_S \text{curl}(\boldsymbol{A}) \text{d} S.
		\end{equation}
	\end{linenomath*}
	where each $\partial S_i$ is a boundary of the surface $S$ and $\text{curl}$ is the 2D curl operator (producing a scalar output).
	This is possible because in 2D, there is only one possible surface bounded by a curve.
	In the 2D equivalent of the infinite-length solenoidal coil, the line integral along a curve $\partial S_1$ enclosing the circle $\partial S_2$ is
	\begin{linenomath*}
		\begin{equation}
			\oint_{\partial S_1} \boldsymbol{A_{is}} \cdot \text{d}\boldsymbol{l}  = \iint_{S_1-S_2} \text{curl}(\boldsymbol{A_{is}}) \text{d}S - \oint_{\partial S_2} \boldsymbol{A} \cdot \text{d}\boldsymbol{l} = 0 + \Phi_B.
		\end{equation}
	\end{linenomath*}
	The curl integral is zero as only the area between the enclosing loop and the boundary of the solenoid is considered where the curl is zero everywhere.
	Along the boundary of the solenoid, the line integral is equivalent to the flux enclosed within the solenoid.
	In 3D, Stokes' theorem thus appears to hold for the case of the infinite-length solenoid, despite being a non-simply connected domain, due to its homotopy equivalence with a circle in 2D.
	
	\subsection{A toroidal solenoid coil}
	\label{sec:torus-theory}

	The magnetic vector potential in the Coulomb gauge outside of a toroidal coil with circular axial and planar cross-sections (\Cref{fig:phase_profiles}B) is closely related to the magnetic field from a simple circular current loop  \cite{carron1995fields}.
	We consider a toroidal solenoid coil with minor radius  $r_s$ and major radius $r_l$ where all the current flows in the poloidal $\hat{\phi}$ direction.
	We introduce the inverse torus aspect ratio
	\begin{linenomath*}
		\begin{equation}
			\zeta = r_s/r_l
		\end{equation}
	\end{linenomath*}
	and shape factor
	\begin{linenomath*}
		\begin{equation}
			g = 2 \frac{1-\sqrt{1-\zeta^2}}{\zeta^2}
		\end{equation}
	\end{linenomath*}
	as the magnitude of the potential fields are dependent on $\zeta$ \cite{carron1995fields}.
	As with the infinite-length solenoidal coil, $\boldsymbol{B}$ is zero everywhere outside the torus volume.
	Inside the torus volume, the magnetic field in cylindrical coordinates is 
	\begin{linenomath*}
		\begin{equation}
			\boldsymbol{B_{t\text{,inside torus volume}}}(r,\theta,z) = \frac{\mu_r \mu_0 N I_t}{2 \pi r} \hat{\theta}
		\end{equation}
	\end{linenomath*}
	where $I_t$ is the current flowing through the wires forming the torus and $N$ is the total number of turns.
	The vector potential outside the toroidal coil volume can be calculated based on the magnetic field $\boldsymbol{B_l}$ from a circular current loop by \cite{carron1995fields}
	\begin{linenomath*}
		\begin{equation}
			\label{eq:torus-A}
			\boldsymbol{A_{t\text{,outside torus volume}}}(r,\theta,z) = r_s  \boldsymbol{B_l}(r,\theta,z)
		\end{equation}
	\end{linenomath*}
	where the corresponding circular current loop has a current
	\begin{linenomath*}
		\begin{equation}
			I_l = \left(\frac{\zeta}{2} \right) N I_t g.
		\end{equation}
	\end{linenomath*}
	A simple circular current loop of radius $r_l$ centered on the $z$-axis and lying in the plane where $z=z_0$ has a magnetic field \cite{schill2003general,smythe1950static}
	\begin{linenomath*}
		\begin{multline}
			\boldsymbol{B_{l}}(r,\theta,z) = \\ \frac{\mu_r \mu_0 I_l}{2 \pi} \frac{(z-z_0)}{r \sqrt{(r+r_l)^2 + (z-z_0)^2}} \left[  -K(k_c) + \frac{r^2 + r_l^2 + (z-z_0)^2}{(r-r_l)^2 + (z-z_0)^2} E(k_c) \right] \hat{r} \\ +
			\frac{\mu_r \mu_0 I_l}{2 \pi}\frac{1}{\sqrt{(r+r_l)^2 + (z-z_0)^2}} \left[  K(k_c) - \frac{r^2 - r_l^2 + (z-z_0)^2}{(r-r_l)^2 + (z-z_0)^2} E(k_c) \right] \hat{z}
		\end{multline}
	\end{linenomath*}
	where
	\begin{linenomath*}
		\begin{equation}
			k_c^2 = \frac{4 r_l r}{(r + r_l)^2 + (z - z_0)^2},
		\end{equation}
	\end{linenomath*}
	and $K(k)$ and $E(k)$ are the complete elliptic integral functions of the first and second kind.
	\begin{linenomath*}
		\begin{equation}
			K(k) = \int_{0}^{\pi/2} \frac{\text{d} \theta'}{(1-k^2 \sin \theta')^{1/2}}
		\end{equation}
	\end{linenomath*}
	\begin{linenomath*}
		\begin{equation}
			E(k) = \int_{0}^{\pi/2} (1-k^2 \sin \theta')^{1/2} \text{d} \theta'.
		\end{equation}
	\end{linenomath*}

	A plot of $\boldsymbol{A_{t,\text{outside torus volume}}}$ and $\boldsymbol{B_{t\text{,inside torus volume}}}$ is shown in \Cref{fig:phase_profiles}E.
	If the current direction were reversed, the direction of the magnetic field and vector potential will switch.
	At the axis of the torus, the potential is vertical.
	The magnitude of the magnetic vector potential outside the torus is strongest near the inner walls of the torus.

	To calculate the phase shift profile for electrons traveling in the $\hat{z}$ ($(r_1,z_0)\rightarrow(r_1,z_1)$) direction, the phase shift profile relative to another electron traveling in the same direction ($(r_0,z_0)\rightarrow(r_0,z_1)$) can be calculated by
	\begin{linenomath*} 
		\begin{multline}
			\label{eq:loop-integral-torus}
			\frac{\hbar}{q_e}\Delta \varphi_r(r_1, \theta,z_0 \rightarrow z_1) = \oint \boldsymbol{A_t}\cdot\text{d}\boldsymbol{l'} \\ =  \int_{r_0}^{r_1} A_{t,r}(r,\theta,z_1) \text{d}r -
			\int_{r_0}^{r_1} A_{t,r}(r,\theta,z_0) \text{d}r  \\ + \int_{z_0}^{z_1} A_{t,z}(r_0,\theta,z) \text{d}z -
			\int_{z_0}^{z_1} A_{t,z}(r_1,\theta,z) \text{d}z.
		\end{multline}
	\end{linenomath*}
	The radial phase shifts from electrons traveling in the positive $\hat{z}$ direction are shown in \Cref{fig:phase_profiles}F and G.
	Unlike the case of conventional electromagnetic lenses which creates phase shifts in both the azimuthal and radial directions, the toroidal lens only causes radial phase shifts.
	Thus the toroidal solenoid coil lens will not induce image rotations.

	Properties of the toroidal solenoid coil electromagnetic lens are compared with conventional electrostatic Einzel and solenoidal electromagnetic lenses in \cref{tab:comparison-table}.
	In general, the toroidal solenoid coil lens takes on the beneficial characteristics of the conventional electrostatic and electromagnetic lenses while having some unique properties of its own.

	\subsection{On the breakdown of Stokes' theorem}
	\label{sec:stokes-breakdown}
	
	Calculations of the lensing properties of the toroidal solenoid coil depend on the breakdown of Stokes' theorem.
	As Stokes' theorem is commonly misapplied in the literature related to the AB effect, we clarify here why it is not applicable.
	Stokes' theorem states that the line integral of a vector field $\boldsymbol{F}$ along the boundary of a surface is equal to the surface integral of the curl of that vector field
	\begin{linenomath*}
		\begin{equation}
			\label{eq:stokes}
			\oint_{\partial S} \boldsymbol{F}\cdot\text{d}\boldsymbol{l} = \iint_S \curl{\boldsymbol{F}}\cdot\text{d}\boldsymbol{S}.
		\end{equation}
	\end{linenomath*}
	Three conditions must be satisfied for this to be true:
	\begin{enumerate}
		\item  $S$ is a piece-wise smooth oriented surface.
		\item $\partial S$ is a simple, closed, piece-wise smooth curve bounding $S$.
		\item The vector field $\boldsymbol{F}$ is continuously differentiable in a region containing $S$.
	\end{enumerate}
	The key difference between Stokes' theorem in 3D and Green's theorem in 2D is that in 3D, there are an infinite number of surfaces that can be bound by a closed curve (\emph{i.e.} the left hand side to right hand side equality of \cref{eq:stokes} must hold for $\emph{any}$ bounded surface and thus $\boldsymbol{F}$ must be continuously differentiable everywhere).
	We have shown that for the case of the infinite-length solenoid, that despite the 3rd condition of continuous differentiability being violated, the equality of Stokes' theorem still holds due to the homotopy equivalence with a 2D system.
	For the 2D torus in 3D space $\mathbb{R}^3 \setminus T^2$, no such homotopy equivalence exists \footnote{Proof: Assume that there exists a set $A \subseteq \mathbb{R}^2$ which is homotopy equivalent to the torus $T^2$. 
		The second singular homology group of the torus is isomorphic to the set of integers $H_2(T^2;\mathbb{Z}) \cong \mathbb{Z}$ \cite{topospaces}. 
		This implies that $H_2(A;\mathbb{Z}) \cong \mathbb{Z}$ as the homology group of any homotopy equivalent sets will be isomorphic. 
		However, as shown by Zastrow \cite{zastrow1999}, $H_n(A;\mathbb{Z}) = 0$ for all $n>1$; leading to a contradiction. 
		Therefore, a homotopy equivalent embedding of the torus in $\mathbb{R}^2$ does not exist.}.
	Therefore, even if a closed loop is formed outside the torus, a surface which crosses the boundary of the torus where the curl is undefined can be formed.

	Through the above analysis of Stokes' theorem, it is clear that the magnetic flux cannot, in general, be equated to a line integral as per \cref{eq:flux} in non-simply connected regions.
	(For non-simply connected spaces, that a vector field is irrotational can not be used to prove that it is also path independent or conservative as can be done in simply connected spaces.)
	Therefore, we define a new quantity: the magnetic circulation along a closed curve $C$ as
	\begin{linenomath*}
		\begin{equation}
			\label{eq:magnetic-circulation}
			\Gamma_B \equiv \oint_C \boldsymbol{A} \cdot \text{d}\boldsymbol{l}.
		\end{equation}
	\end{linenomath*}
	It is then generally true that the AB effect is proportional to the magnetic circulation enclosed by two charged particle paths and not the enclosed magnetic flux
	\begin{linenomath*}
		\begin{equation}
			\Delta \varphi_\text{AB} = \frac{q}{\hbar} \Gamma_B \neq \frac{q}{\hbar} \Phi_B.
		\end{equation}
	\end{linenomath*}

	\subsection{Stationary action methods}
	\label{sec:lagrangian}

\begin{figure}[]
		\centering
		\includegraphics[width=1\linewidth]{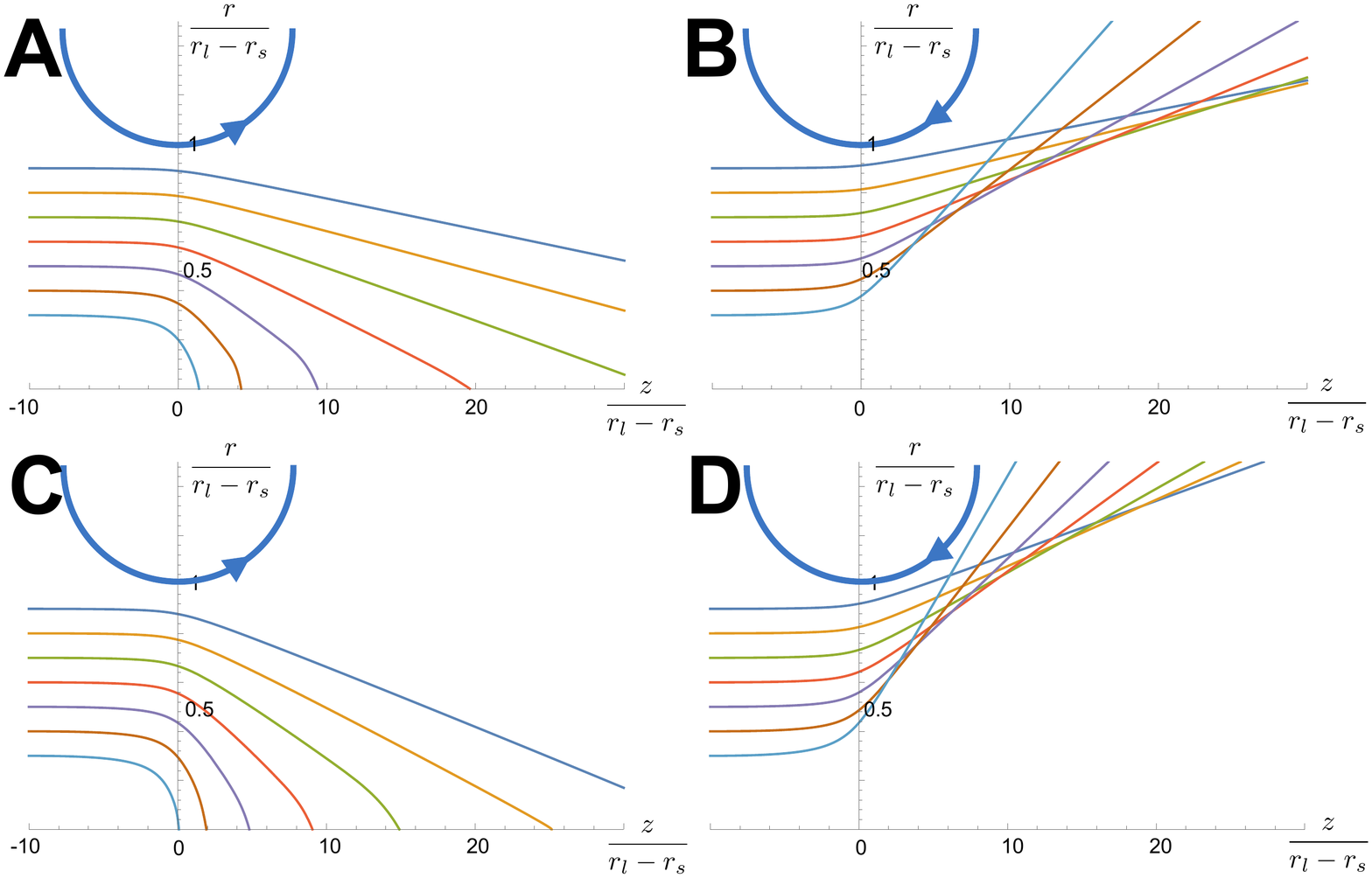}
		\caption{\textbf{Calculation of the classical ray trajectories for 	parallel electron rays entering inside the hole of a toroidal solenoid coil.} The blue semicircles indicate the walls of the torus. (\textbf{A}, \textbf{C}) and (\textbf{B}, \textbf{D}) are opposite current-flow directions as indicated by the arrow on the torus wall. (\textbf{A},\textbf{B}) are calculations for 300\,keV electrons and (\textbf{C}, \textbf{D}) are for 100\,keV electrons.  The horizontal axis is the optical ($z$) axis and the vertical axis is the radial motion. Each colored line represents a different electron trajectory.}
		\label{fig:torus-ray-trajectories}
	\end{figure}

	From a classical geometrical optics perspective, lensing effects should be describable by the principle of stationary action.
	We start with Fermat's principle
	\begin{linenomath*}
		\begin{equation}
			\label{eq:fermat-stationary-action}
			\delta \int_{\boldsymbol{x}_0}^{\boldsymbol{x}_f} n \text{d}s = 0
		\end{equation}
	\end{linenomath*}
	where $\delta$ is the variational, $\boldsymbol{x}_0$ and $\boldsymbol{x}_f$ are the initial and final positions of the particle traversing a path $s$, and $n$ is the electron-optical refractive index \cite{groves2014charged}
	\begin{linenomath*}
		\begin{equation}
			n = p +q \boldsymbol{A} \cdot \boldsymbol{\hat{s}}
		\end{equation}
	\end{linenomath*}
	where $p$ is the kinematic momentum magnitude of the particle.
	For axially symmetric systems, we can parameterize the particle motion in terms of the $z$(optic)-axis to arrive at a reduced refractive index in cylindrical coordinates of
	\begin{linenomath*}
		\begin{equation}
			\label{eq:reduced-refractive}
			n_\text{red}(r,\theta,z) = p \sqrt{1+ r'^2 + r^2 \theta'^2} +q(r' A_r + r \theta' A_\theta + A_z)
		\end{equation}
	\end{linenomath*}
	where the primes represent differentiation with respect to $z$ and $(A_r,A_\theta,A_z)$ are the cylindrical components of the magnetic vector potential.
	The paths $\boldsymbol{r}=(r(z),\theta(z))$ of a particle passing through the lens can then be solved through the eikonal equation
	\begin{linenomath*}
		\begin{equation}
			\label{eq:eikonal}
			\frac{\partial n_\text{red}}{\partial \boldsymbol{r}} - \frac{\text{d}}{\text{d}z}\frac{\partial n_\text{red}}{\partial \boldsymbol{r'}} = 0.
		\end{equation}
	\end{linenomath*}
	For the case of the toroidal solenoid lens, electric potential distribution will be treated as uniform such that the kinematic momentum magnitude is a constant $p_0$.
	As there is no azimuthal component to the magnetic vector potential generated by the toroidal solenoid, no rotational motion is induced by the lens in contrast to conventional short solenoidal coil lenses.
	The radial equation of motion simplifies to
	\begin{linenomath*}
		\begin{equation}
			\label{eq:fermat-torus}
			\frac{p_0 r''}{(1+r'^2)^{3/2}}=q\left(r'\frac{\partial A_r}{\partial r} \right).
		\end{equation}
	\end{linenomath*}
	Note that this is only valid in the regions where the magnetic field is zero.
	The radial motion is solved numerically for parallel rays entering the central hole of a toroidal solenoid coil in \cref{fig:torus-ray-trajectories} for two different accelerating voltages at equal and opposite current flows.
	In \cref{fig:torus-ray-trajectories}A and C, convex lensing with a negative spherical aberration is shown.
	In \cref{fig:torus-ray-trajectories}B and D, concave lensing with a positive spherical aberration is shown.
	Thus the particle trajectories calculated through this geometrical optics formulation are consistent with the motions predicted through the phase-shift analysis.
	Comparing the changes in trajectory of \cref{fig:torus-ray-trajectories}A and B at 300\,keV with \cref{fig:torus-ray-trajectories}C and D, at 100\,keV, it is also clear that the chromatic aberration coefficient is of the same polarity as the focal polarity.

	\section{Experimental}
	
	\subsection{Holography experiments}
	\label{sec:holography}
	
\begin{figure}
		\centering
		\includegraphics[width=1\linewidth]{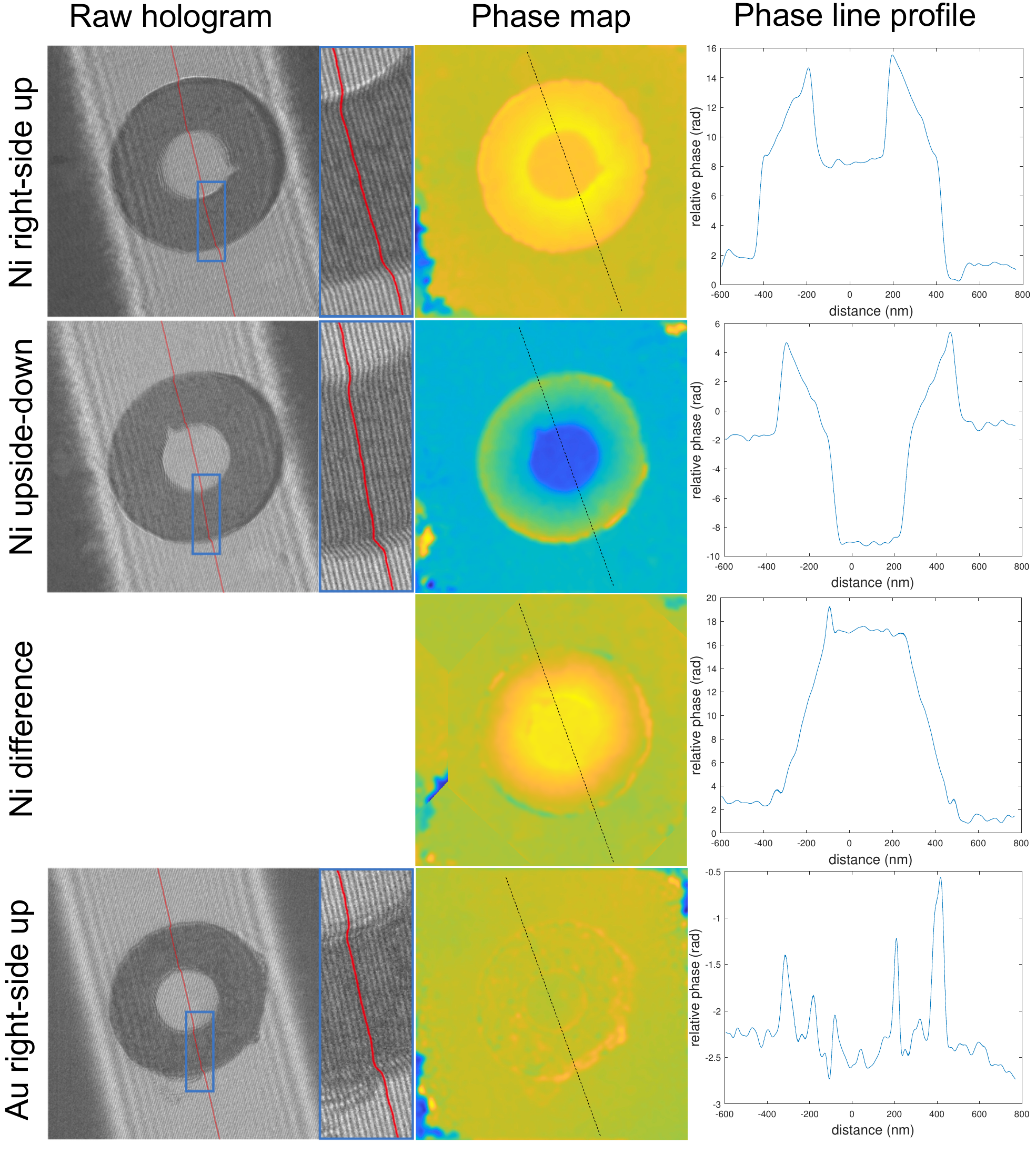}
		\caption{\textbf{Examples of electron holograms of nanoring samples.}  Left-hand column: Raw holograms. A single fringe of the holograms is highlighted in red. The area in the blue box is magnified on the right-hand side of the image.
			Middle column: Reconstructed phase maps.
			Right-hand column: Line profiles through the phase maps. The 	location that the line profile is measured from is indicated with the dashed line on the phase maps.
			Top row: Magnetic nickel nanoring in the right-side up orientation.
			Second row: Same magnetic nickel nanoring as above but in the 	upside-down orientation.
			Third row: Phase difference map of the upside-down Ni nanoring from 	the right-side up Ni nanoring (after rigid rotational and translational alignment) to remove the electrical contribution to the phase map.
			Last row: Non-magnetic gold nanoring of similar dimension to the 	nickel nanoring.}
		\label{fig:holograms}
	\end{figure}

	The magnetization states of the nanoring samples were measured using electron holography.
	Holograms were obtained for both magnetic nickel and non-magnetic gold nanorings in both right-side up and upside-down orientations.
	An example of the obtained holograms are shown in \cref{fig:holograms}.
	The holograms exhibit phase shift contributions from both electrical and magnetic effects.
	The electrical contributions are from the mean inner potential of the sample and substrate as well as any beam-induced charging effects.
	A red line highlights a single fringe in the raw holograms on the left-hand side of the figure.
	Here, a sudden lateral shift in the fringe position can be observed between the substrate and nanoring transition boundaries in both the magnetic and non-magnetic nanorings.
	This is due the the mean inner potential of the nanoring material.
	For the magnetic nanorings, an angle exists between the fringe inside and outside the nanoring volume.
	This is due to the magnetic contribution.
	For the non-magnetic gold nanorings, the fringes in the volume of the nanorings are parallel to the fringes outside the nanorings as there are no magnetic contributions.
	In the reconstructed phase maps (middle column of the figure), a large phase jump can be observed between the substrate and nanoring transition boundaries of the magnetic nanorings.
	This jump should occur for the gold nanorings but the phase-unwrapping procedure could not be performed successfully possibly due to local reductions in fringe contrast within the denser and higher atomic mass gold.
	It is clear from the fringes in the raw hologram that this phase jump is present.
	Inside the volume of the nickel nanorings, a gradual phase ramp is observed due to the vortex magnetization state.
	These two contributions are clear in the line-profiles through the nanorings in the right-hand side of the figure.
	We note that the magnetic nanorings used here have a rectangular axial cross-section as opposed to the circular axial cross section in the calculated toroidal solenoid coils. 
	The rectangular axial cross section makes the magnetic phase ramp more linear than the phase ramp in \Cref{fig:phase_profiles}F.
	In the third row of \cref{fig:holograms}, a difference map between the phase reconstructions of the right-side up and upside-down ring orientations after rigid translational and rotational alignments is shown.
	This subtraction is performed to remove the electrical phase shift contribution and double the magnetic contribution.
	The difference map clearly demonstrates the large phase shift step-change between the regions outside and inside the hole of the nanorings as expected by the traditional understanding of the AB effect.
	Such a phase map is only possible for nanorings with a vortex magnetization state.
	This change in phase between the inner hole and area outside the nanoring is absent in the non-magnetic gold nanorings as is expected.

	\subsection{Fresnel diffraction experiments}
	\label{sec:fresnel-diff}
	
	The position of Fresnel fringes are sensitive to phase objects in-between the source and detector.
	We utilize this to demonstrate the lensing effect through toroidal magnetic fields on samples whose magnetization states were confirmed through electron holography.
	In principle, for a magnetic nanoring containing a strong enough magnetic flux, a change in the image focus between inside and outside the nanoring should be observable.
	However,  the low signal to noise from image measurements make it highly challenging to detect the minute focal length changes expected through the magnetic nanoring structures. 
	Fresnel diffraction measurements provide higher signal as all the electron beam intensity is concentrated -- making possible the observation of very small focal-length changes.

	Fresnel diffraction patterns from a convergent electron beam focused inside the nanorings were measured and compared with Fresnel patterns obtained with the same beam in a vacuum area far from the nanorings.
	Vacuum was used instead of adjacent substrate to eliminate the effects of small variations in charge-state and thickness across the support membrane.
	Examples of the Fresnel patterns are in \Cref{fig:diffraction_fresnel_expt}.
	In the experimental diffraction patterns obtained, some reductions in fringe intensity were observed inside the nanorings for both the magnetic and non-magnetic samples.
	For the non-magnetic gold nanorings, the position and spacing of the Fresnel fringes were practically unchanged from inside and outside the nanorings.
	For magnetic nickel nanorings with thicker walls, larger shifts of the Fresnel fringes inside the nanoring relative to the fringes in vacuum were observed.
	When the nanoring orientation was flipped, the direction that the fringes shifted was in the opposite manner.

	\begin{figure}
		\centering
		\includegraphics[width=0.7\linewidth]{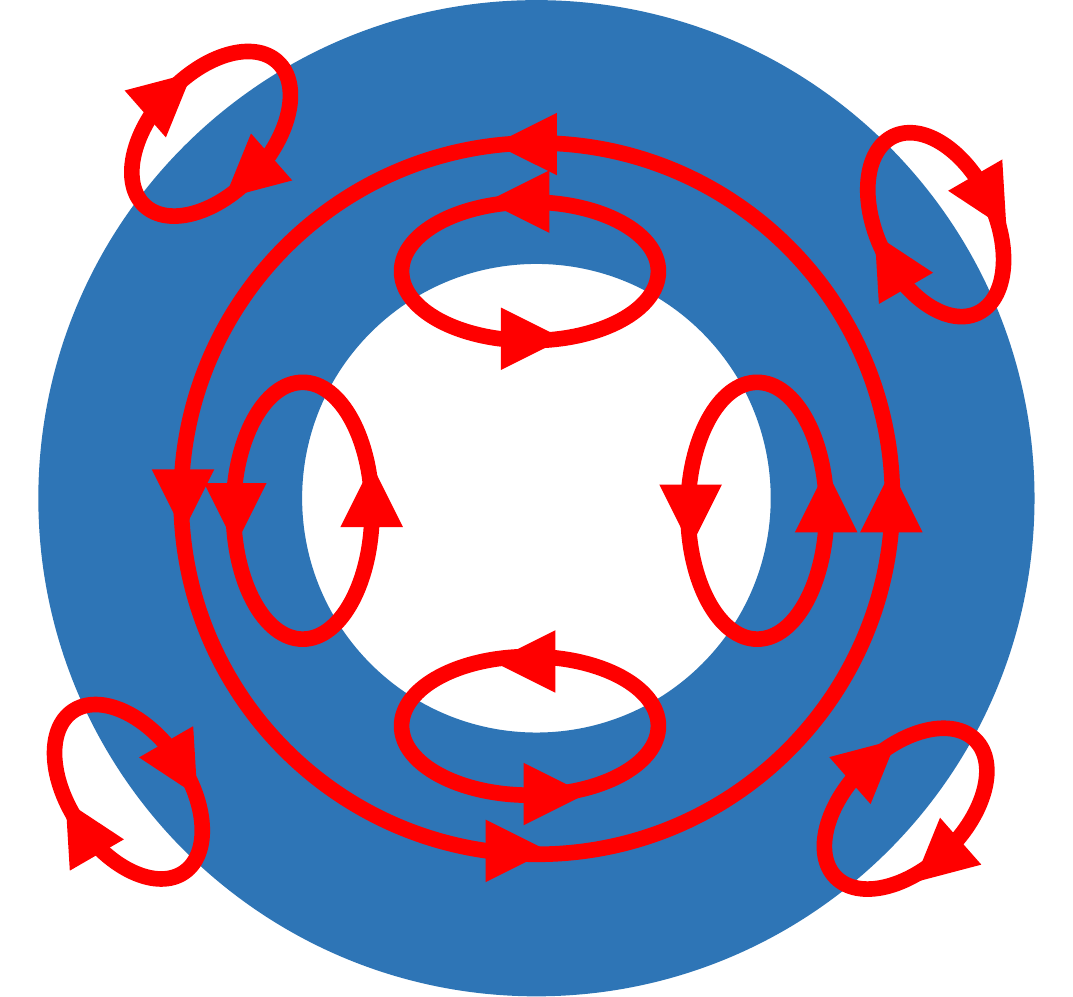}
		\caption{Diagram representing possible stray magnetic fields (red loops) in a non-axially symmetric toroidal solenoid coil (blue ring).}
		\label{fig:torus-fringing-fields}
	\end{figure}	
	
	\begin{figure}
		\centering
		\includegraphics[width=0.5\linewidth]{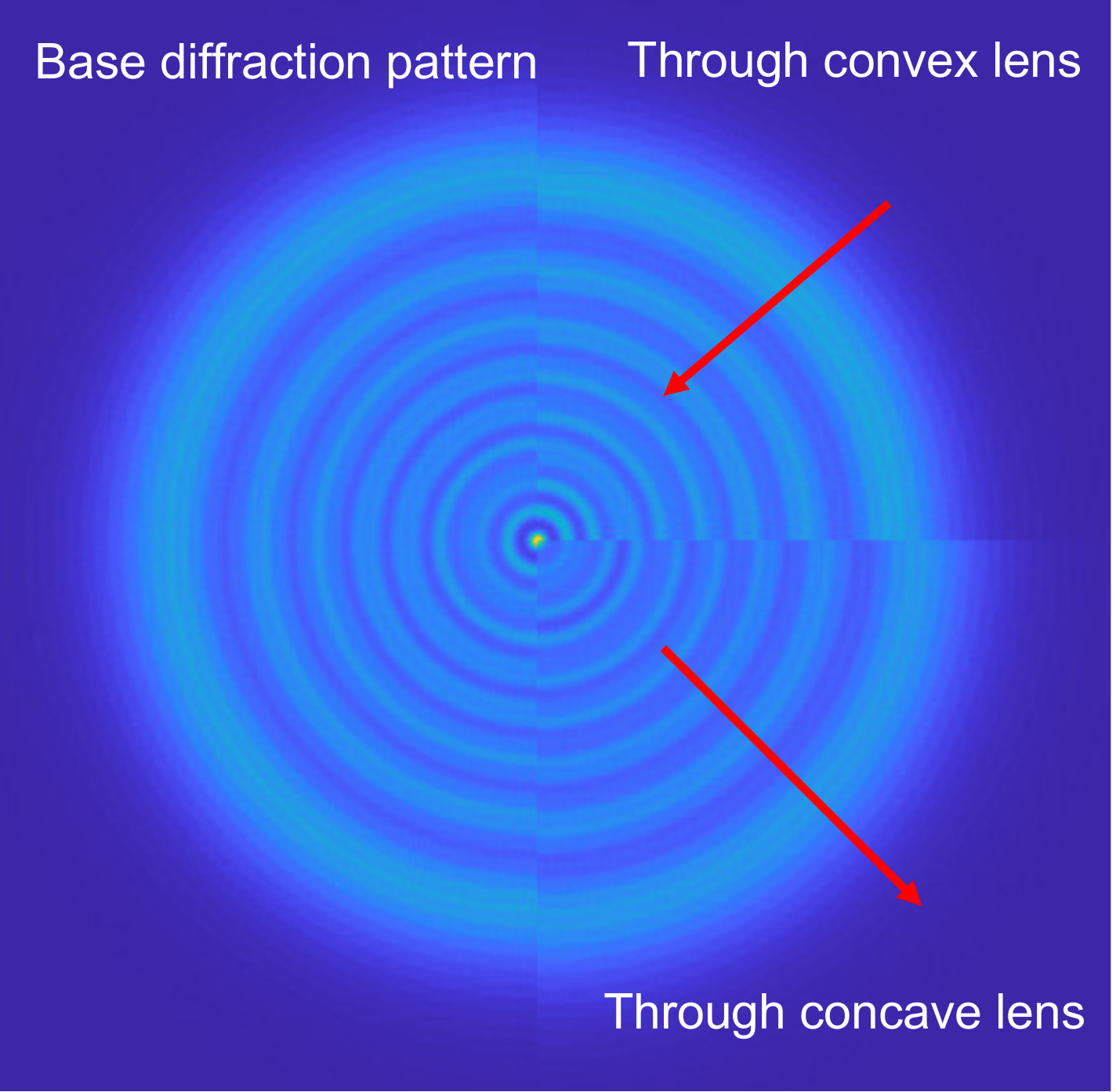}
		\caption{\textbf{Simulations of Fresnel diffraction from an aperture compared to diffraction through a convex and concave lens with equal and opposite focal lengths.} The arrows represent the direction the Fresnel fringes of the lensed patterns shift with respect to the non-lensed diffraction pattern. Fresnel patterns were simulated by standard Fresnel propagation algorithms \cite{voelz2011computational} from an ideal lens where the phase function at the detection plane is $\varphi(r) = \frac{\pi r^2}{\lambda z} -\frac{\pi r^2}{\lambda f}$	where $\lambda$ is the wavelength, $z$ is the propagation distance, and $f$ is the focal length.}
		\label{fig:fresnel_diff_lens_sim}
	\end{figure}
	
	\begin{table}[]
		\caption{Comparison of the properties of conventional charged particle lenses with those of the toroidal solenoid coil lens proposed in the present paper.}		\includegraphics[width=1\linewidth]{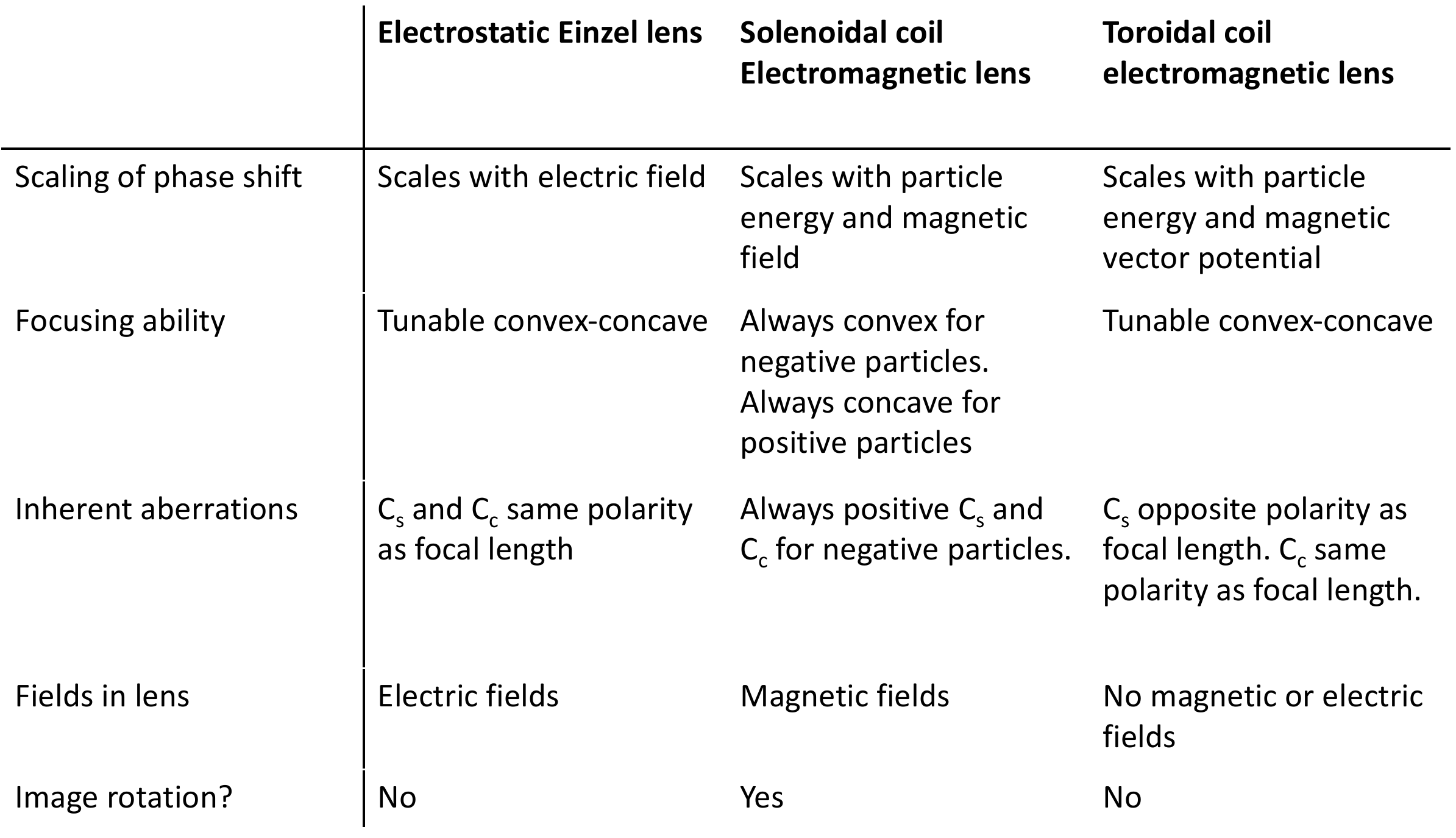}
		\label{tab:comparison-table}
	\end{table}
	
	\clearpage

\end{document}